\documentclass[conference,a4paper]{APSIPA2025}
\usepackage{amsmath}
\usepackage{graphicx}
\usepackage{multirow}
\usepackage{amssymb}
\usepackage{bm}
\usepackage{threeparttable}
\usepackage[backend=biber,style=ieee,]{biblatex}
\usepackage{geometry}
\usepackage{fancyhdr}

\fancypagestyle{firststyle}{
  \fancyhf{}
  \fancyhead[C]{2025 Asia Pacific Signal and Information Processing Association Annual Summit and Conference (APSIPA ASC)}
}

\begin{document}

\title{A Robust Proactive Communication Strategy for Distributed Active Noise Control Systems}

\author{ 
\authorblockN{
Junwei Ji\authorrefmark{1}, Dongyuan Shi\authorrefmark{2}, Zhengding Luo\authorrefmark{1}, Boxiang Wang\authorrefmark{1}, Ziyi Yang\authorrefmark{1}, Haowen Li\authorrefmark{1}, Woon-Seng Gan\authorrefmark{1}
}

\authorblockA{
\authorrefmark{1}
Nanyang Technological University, Singapore\\
Email: \{JUNWEI002, luoz0021, boxiang001, ziyi016\}@e.ntu.edu.sg, \{haowen.li, ewsgan\}@ntu.edu.sg}

\authorblockA{
\authorrefmark{2}
Northwestern Polytechnical University, Xi'an, China\\
Email: dongyuan.shi@nwpu.edu.cn}
}

\maketitle
\thispagestyle{firststyle}
\pagestyle{fancy}

\begin{abstract}
  Distributed multichannel active noise control (DMCANC) systems assign the high computational load of conventional centralized algorithms across multiple processing nodes, leveraging inter-node communication to collaboratively suppress unwanted noise. However, communication overhead can undermine algorithmic stability and degrade overall performance. To address this challenge, we propose a robust communication framework that integrates adaptive-fixed-filter switching and the mixed-gradient combination strategy. In this approach, each node independently executes a single-channel filtered reference least mean square (FxLMS) algorithm while monitoring real-time noise reduction levels. When the current noise reduction performance degrades compared to the previous state, the node halts its adaptive algorithm, switches to a fixed filter, and simultaneously initiates a communication request. The exchanged information comprises the difference between the current control filter and the filter at the time of the last communication, equivalent to the accumulated gradient sum during non-communication intervals. Upon receiving neighboring cumulative gradients, the node employs a mixed-gradient combination method to update its control filter, subsequently reverting to the adaptive mode. This proactive communication strategy and adaptive-fixed switching mechanism ensure system robustness by mitigating instability risks caused by communication issues. Simulations demonstrate that the proposed method achieves noise reduction performance comparable to centralized algorithms while maintaining stability under communication constraints, highlighting its practical applicability in real-world distributed ANC scenarios.
\end{abstract}

\section{Introduction}
Noise issue has attracted more and more attention in recent years, since it not only causes some hearing problems but also has negative effects on the cardiovascular and metabolic systems and even cognitive impairment problems \cite{peris2020noise}. Passive noise control, which blocks noise propagation, is effective for high-frequency noise, while the active noise control (ANC) technique has better performance for low-frequency noise, such as compressor and engine noise \cite{Kuo1999ANC}. The principle behind the ANC is the wave superposition strategy \cite{lueg1936process}. It generates the anti-noise that has the same amplitude but opposite phase to the noise and then takes effect on the acoustic field. Due to the variation of the noise and the acoustic environment, the adaptive algorithm is considered. The filtered reference least mean square (FxLMS) algorithm \cite{Morgan1980FXLMS} is one of the most widely used algorithms in the ANC field. Recently, some novel algorithms based on the FxLMS algorithm have been proposed to improve the performance on the practical issues \cite{Kajikaea2012ANC,Lam2021Ten,shen2024survey,guo2024survey,moreau2008review,wang2025transferable,george2013advances}. With the development of artificial intelligence (AI) techniques, some deep learning based ANC methods have sprung up \cite{zhang2021deep,Luo2023GFANC,Luo2024ImplemSFANC,xie2024cognitive} to further improve the system's stability and noise reduction performance.


In recent years, there has been a growing interest in achieving large-area global noise reduction. This trend has spurred the development and drawn increased attention to multichannel active noise control (MCANC) systems. These systems deploy multiple loudspeakers and microphones to attenuate unwanted noise \cite{iwai2019multichannel,lorente2014gpu}.  The conventional centralized strategy requires a single processor to not only generate control signals but also update control filters by collecting all inputs. One of the typical algorithms is the multiple error FxLMS (MEFxLMS) algorithm \cite{Elliott1987MEANC}. Therefore, it places a substantial demand on processor performance to process its massive computational cost. To reduce the computational complexity of this kind of centralized strategy, several efficient centralized algorithms have also been proposed \cite{murao2017mixed,shi2021block}. Alternatively, decentralized strategies distribute computation across multiple controllers, each updating its filter using only its local error signals \cite{Zhang2019Decentralized, george2012particle}. However, it typically overlooks inter-node acoustic crosstalk, which can compromise stability and performance in practice \cite{Kuo1999ANC}. 

Hence, the distributed MCANC (DMCANC) system is developed to balance the advantages of centralized and decentralized control strategies. DMCANC system consists of several ANC nodes, where each node processes its own signals independently, while exchanging certain information with other nodes to ensure global noise reduction performance \cite{Ferrer2015DistributedANC,Ferrer2017Distributed}. The diffusion strategy \cite{Chen2022DistributedANC,Song2016DiffusionANC} is widely considered in the DMCANC system compared to the incremental strategy \cite{Ferrer2015DistributedANC}, since it only cooperates with nodes' neighbors, resulting in low communication requirements. The conventional diffusion FxLMS (DFxLMS) algorithm generates the global control filter from other nodes' local control filters through topology-based combination rules \cite{Chu2020DiffusionANC,Chu2021Combination}. Recently, the augmented DFxLMS (ADFxLMS) has been proposed to improve the performance on asymmetric paths \cite{Li2023Distributed,Li2023AugmentedDiffusion}, and the auto-shrink step size mixed-gradients distributed FxLMS (ASSS-MGDFxLMS) is developed to prevent the system's instability from communication delay in the distributed network \cite{ji2025mixed}. However, frequent communication also places an additional burden on the system.

Therefore, a proactive communication strategy for the DMCANC systems (PC-DMCANC) is introduced. In our scheme, each node independently conducts a single‐channel FxLMS algorithm while continuously evaluating its noise‐reduction performance. Whenever a node detects a degradation in attenuation relative to its previous state, it temporarily suspends adaptation, reverts to a fixed filter mode, and issues a communication request to its neighbors. During the non‐communication interval, the node accumulates the incremental gradients—i.e., the differences between its current control filter and the filter at the time of the last communication. Upon receiving neighboring cumulative gradients, the node employs a mixed-gradient combination method to update its control filter, then resumes adaptive operation. This proactive communication strategy and adaptive-fixed switching mechanism ensure system robustness by mitigating instability risks caused by communication issues, underscoring its suitability for real‐world distributed ANC applications.

The remainder of this paper is organized as follows: Section \ref{sec:method} begins with a brief overview of DMCANC, followed by our proposed PC-DMCANC method. Section \ref{sec:sim} evaluates the performance of the algorithm through numerical simulation studies. Finally, the conclusion is drawn in Section \ref{sec:conclusion}.

\section{Methodology}\label{sec:method}

\subsection{Distributed Multichannel ANC system}

\begin{figure}[!t]
    \centering
    \includegraphics[width=0.9\columnwidth]{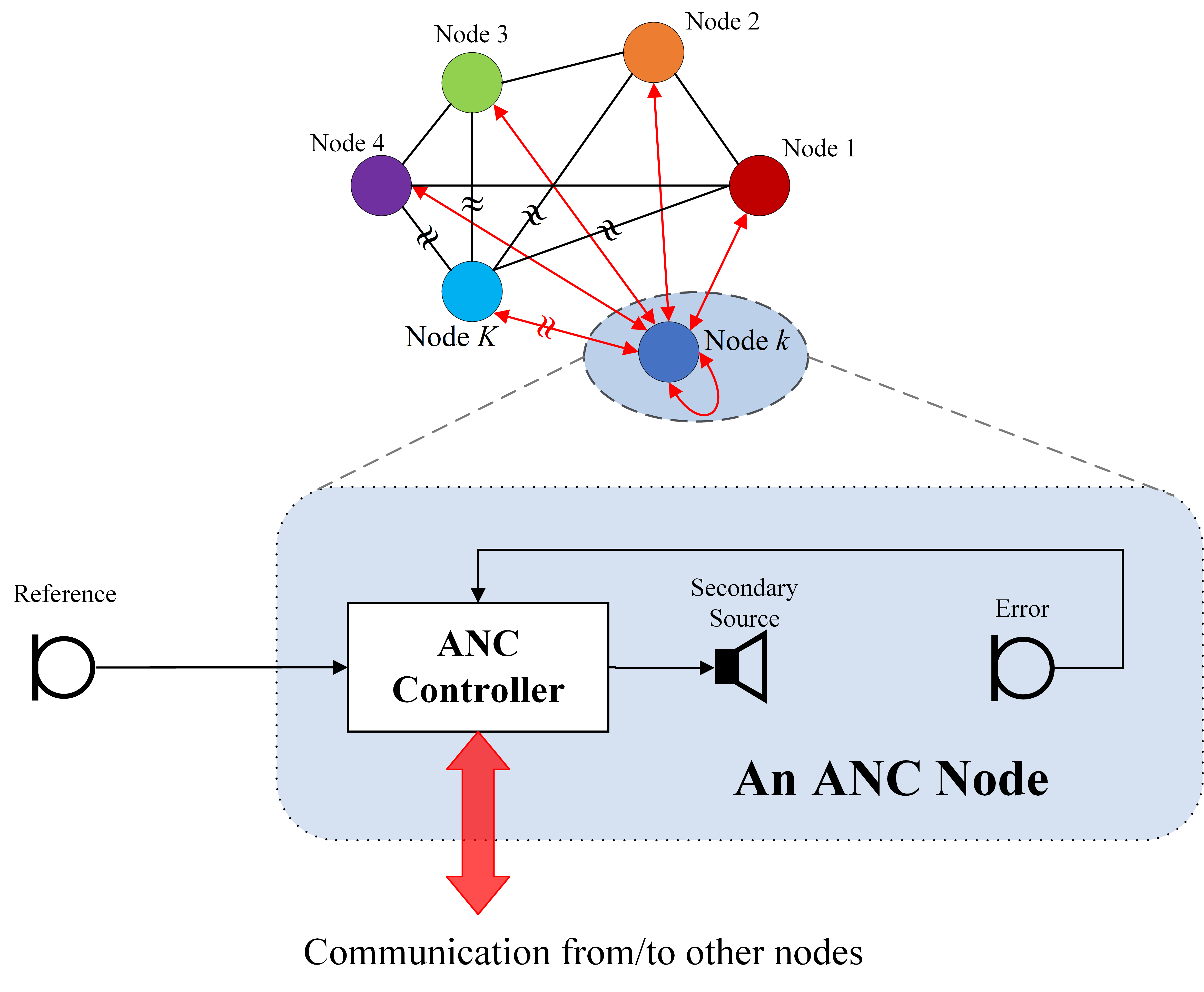}
    \caption{A DMCANC network, where each ANC node consists of a secondary source, an error microphone, and an ANC controller, and shares with one reference microphone.}
    \label{fig:ancnode}
\end{figure}

The multichannel ANC (MCANC) system is widely used to create a large zone of quiet (ZoQ). Due to the high computational cost of the conventional centralized strategy, the distributed MCANC (DMCANC) methods are involved in the MCANC system to improve the computational efficiency. It distributes huge computing tasks among several ANC nodes as shown in Fig.~\ref{fig:ancnode}, where each node is composed of a secondary source, an error microphone, and an ANC controller for signal processing and information exchange with other nodes. Figure~\ref {fig:distributed} illustrates a DMCANC system with $K$ nodes, where the $k$th node generates the control signal as:
\begin{equation}\label{eq1:controlsignal}
     y_k(n) = \mathbf{w}_k^\mathrm{T}(n)\mathbf{x}(n), \quad k = 1,2,...,K,
\end{equation}
in which $\mathbf{x}(n)=[x(n) \, x(n-1) \, \cdots \, x(n-N+1)]^\mathrm{T}$ denotes the reference signal vector, $\mathbf{w}_k(n)=[w_{k,1}(n) \, w_{k,2}(n) \, \cdots \, w_{k,N}(n)]^\mathrm{T}$ stands for the control filter with the tap length of $N$, and $n$ is the time index. Hence, the residual error signal at the $k$th node can be expressed as:
\begin{equation}\label{eq2:errorsignal}
     e_k(n) = d_k(n) - y_k(n)*s_{kk}(n)-\sum_{m=1,\neq k}^{K}y_m(n)*s_{km}(n),
\end{equation}
where $*$ denotes the linear convolution, $d_k(n)$ represents the disturbance signal, ${s}_{kk}(n)$ and ${s}_{km}(n)$ refers to the impulse response of node's self-secondary path and cross-secondary paths between the nodes from the $m (m\neq k)$th secondary source to the $k$th error sensor, respectively. 

The conventional DMCANC system has two primary processes: adaptation and combination. During the adaption, each node minimizes its local error signal using an FxLMS-based algorithm. Then, some essential information, such as control filters or gradients, will be exchanged in the distributed network, followed by the combination operation to ensure the stability and satisfactory noise reduction performance. In the combination phase, the topology-based rule \cite{Chu2021Combination} is widely used. However, this approach ignores the actual acoustic path effects and is particularly ineffective for asymmetric paths and inter-node acoustic crosstalk effects. It is obvious from \eqref{eq2:errorsignal} that the last term can be regarded as the interference from other nodes, which can be expressed as:
\begin{figure}[!t]
    \centering
    \includegraphics[width=0.9\columnwidth]{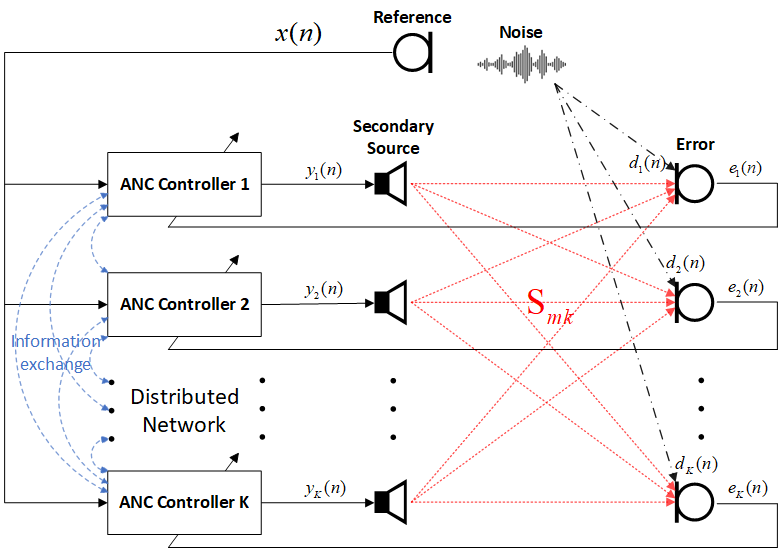}
    \caption{The schematic diagram of DMCANC, where each ANC controller exchanges information through a distributed network \cite{ji2025mixed}.}
    \label{fig:distributed}
\end{figure}
\begin{equation}\label{eq3:interference}
    \gamma_k(n) = \sum_{m=1,\neq k}^{K}y_m(n)*s_{km}(n).
\end{equation}

Another combination approach is to introduce compensation filters, $c_{km}(n)$, which are used to make up for the difference between the self-secondary path and cross-secondary paths \cite{ji2025mixed}, described as:
\begin{equation}\label{eq4:compensate}
 s_{km}(n) = s_{kk}(n) * c_{km}(n), \; (m\neq k).
\end{equation}
The compensation filter-based strategy includes the acoustic information during the combination stage, resulting in overcoming the cross-talk effect effectively. Therefore, the compensation filter-based method is considered in this paper.

\subsection{Mixed Cumulative Gradient (MCGD) technique}
\begin{figure}[!t]
    \centering
    \includegraphics[width=0.9\columnwidth]{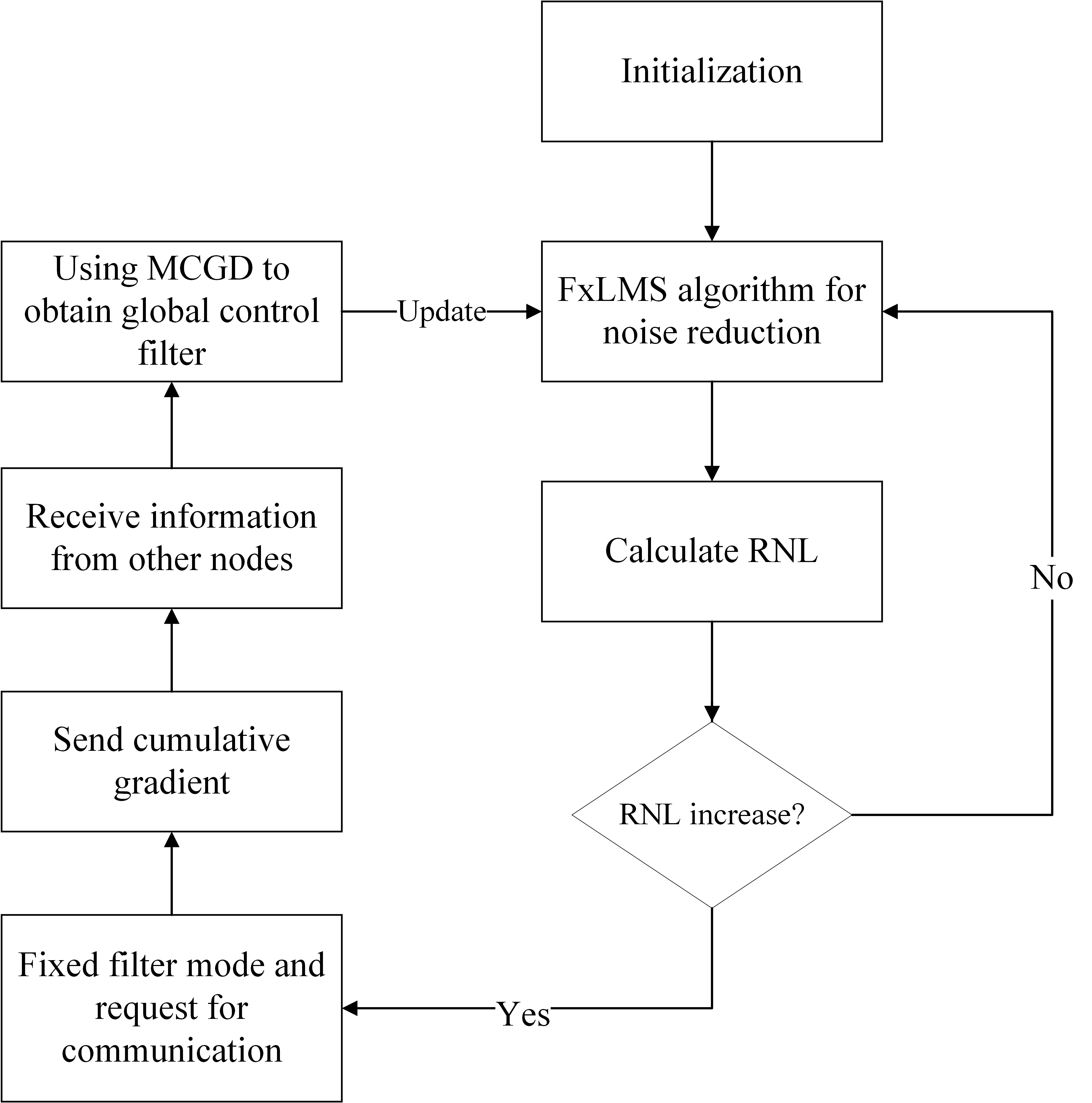}
    \caption{The flow chart of the proposed PC-DMCANC for the $k$th node.}
    \label{fig:flow}
\end{figure}
In the DMCANC system, each node uses its own error signal as feedback to update the control filter through a single-channel FxLMS algorithm. The update equation should be:
\begin{equation}\label{eq5:adaptation}
    \mathbf{w}_k(n+1) = \mathbf{w}_k(n) + \mu_{k} \mathbf{x}_{kk}'(n)e_k(n),
\end{equation}
where $\mathbf{x}_{kk}'(n)$ denotes the filtered reference signal vector as
\begin{equation}\label{eq6:filteredx}
    \mathbf{x}_{kk}'(n) = \hat{s}_{kk}(n)*\mathbf{x}(n).
\end{equation}
To guarantee the noise reduction performance, data needs to be exchanged between nodes through a distributed network. However, ANC requires high real-time performance; hence, most DMCANC algorithms assume that the communication network is ideal, i.e., every sampling point can complete the communication task, which is unrealistic. On the other hand, a high communication frequency will bring an extra burden to the system. Therefore, we try to reduce the communication frequency. 

In order to allow the DMCANC system to acquire valid data to ensure convergence of the algorithm, the differences between the current control filter and the filter at the time of the last communication will be transmitted.

Assuming that the control filter at the last communication is denoted as $\mathbf{w}_k'$ and the transmitted difference is defined as:
\begin{equation}\label{eq7:differ}
    \boldsymbol{\phi}_k(n) = \mathbf{w}_k(n) - \mathbf{w}_k'.
\end{equation}
Equation \eqref{eq7:differ} can also be regarded as the cumulative gradient during the non-communication phase. If the $k$th node receives other nodes' transmitted cumulative gradients, the combination will be executed as:
\begin{equation}\label{eq8:combination}
    \mathbf{w}_{k}^{new}(n) = \mathbf{w}_k' + \boldsymbol{\phi}_k(n) + \sum_{m=1,m\neq k}^K \boldsymbol{\phi}_m(n)*c_{km}(n).
\end{equation}
The newest global control filter $\mathbf{w}_k^{new}(n)$ for the $k$th node, obtained through the mixed cumulative gradient (MCGD) technique, is then updated to real controller $\mathbf{w}_k(n)$ in the system for noise cancellation and subsequent FxLMS updates. 

\subsection{Proactive communication strategy}
\begin{figure}[!t]
    \centering
    \includegraphics[width=0.9\columnwidth]{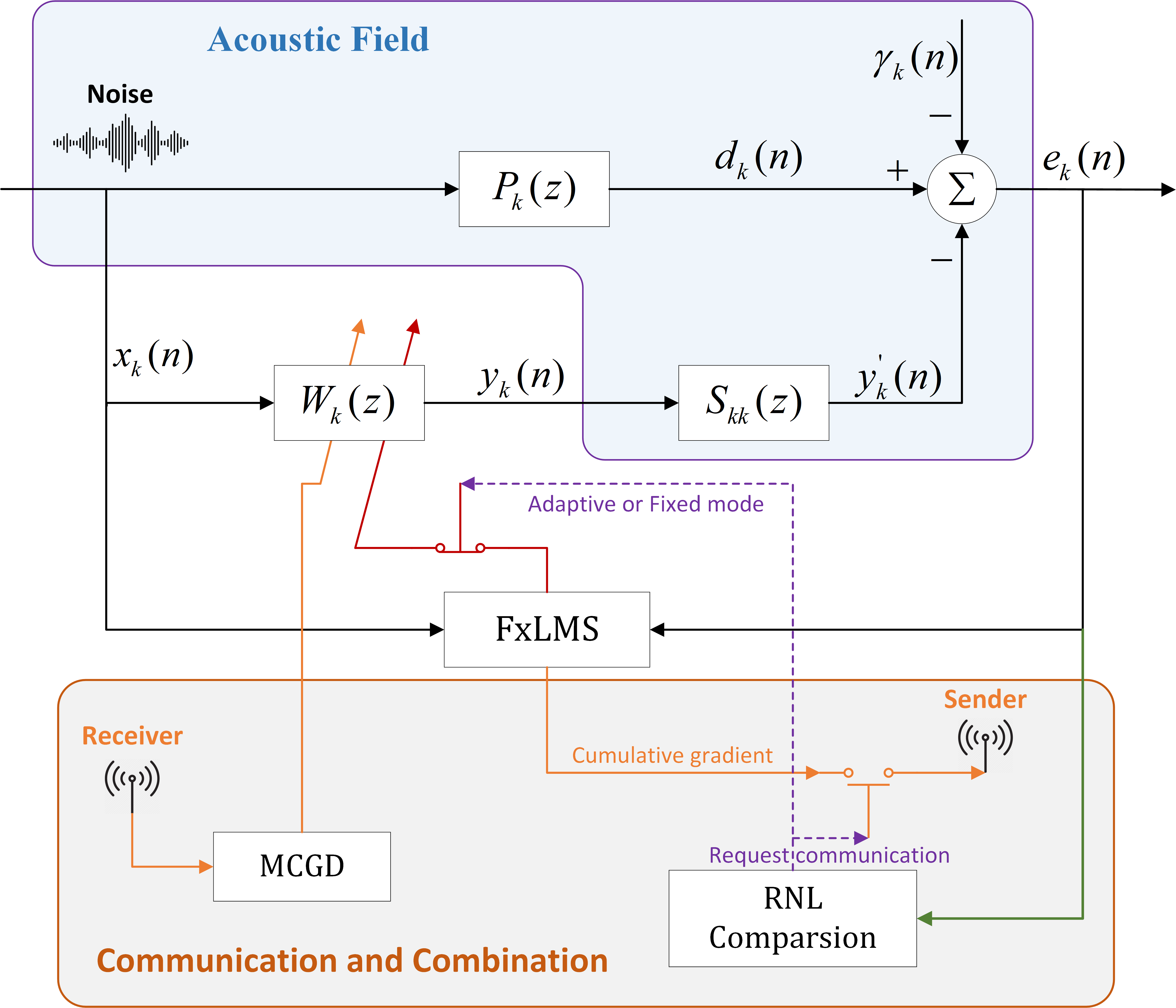}
    \caption{The block diagram of the proposed PC-DMCANC for the $k$th node, where $P_k(z)$ and $S_{kk}(z)$ represent the primary path and self-secondary path, respectively.}
    \label{fig:pcdmanc}
\end{figure}
As mentioned before, conventional DMCANC systems require real-time communication, which is greatly affected by the communication environment and thus causes system instability. Therefore, we introduce a proactive communication strategy, where nodes determine the time for information exchange. The brief flow chart of the proposed proactive communication DMCANC (PC-DMCANC) system for each node is illustrated in Fig.~\ref{fig:flow}.

Initially, each ANC node eliminates the noise using a single-channel FxLMS algorithm while monitoring the noise reduction performance. Therefore, the average residual noise level (RNL) of a certain frame is considered to evaluate the node's performance, which is defined as:
\begin{equation}\label{eq9:anrl}
    \bar{\eta}_k = 10\times\log10\{\frac{1}{Tf}\sum_{t=n-Tf}^{n}\left[{e^2_k(t)}\right]\},
\end{equation}
where $f$ denotes the sampling frequency in Hz and $T$ represents the frame period in seconds. When the RNL value increases, indicating that the system may diverge or fail to converge further, the node changes its adaptive mode to a fixed filter mode to maintain noise reduction performance, simultaneously requesting communication and sending the cumulative gradient defined in Eq.~\eqref{eq7:differ}. When other nodes receive a communication request from a node, they also switch to fixed filter mode while sending their accumulated gradients. Subsequently, all the nodes simultaneously use the MCGD method to get the newest global control filter and apply it to the system. Finally, each node switches back to adaptive mode to further reduce the external noise and so on. In contrast to the traditional DMCANC, which requires communication at each sampling point, the proposed method allows each node to initiate communication spontaneously based on its own noise reduction level to obtain global information.

A detailed block diagram of the proposed PC-DMCANC algorithm for each node is depicted in Fig.~\ref{fig:pcdmanc}. The MCGD method can alleviate the acoustic crosstalk effect between nodes by integrating the cumulative gradient of each node using compensating filters, thus achieving satisfactory global noise reduction. The proactive communication strategy effectively reduces the communication frequency. The adaptive-fixed-filter switching mechanism avoids the instability caused by communication delays and further divergence during the communication phase. Besides, the computational complexity is also dispersed in space and time.

\section{Numeric simulations}\label{sec:sim}
In this section, the performance of our proposed PC-DMCANC approach is validated in an MCANC system with 6 ANC nodes. The primary and secondary paths are measured on an ANC window with the same configuration as \cite{ji2025mixed}. The control filter and secondary path are modeled with tap lengths of 512 and 256, respectively, and the sampling frequency is 16,000Hz. The residual noise level (RNL) is computed with the frame period of 0.1 seconds, resulting in at least $10\%$ communication frequency reduction. The average normalized squared error (ANSE) across all ANC nodes is applied to evaluate the noise reduction (NR) performance of the algorithms, which is defined as
\begin{equation}\label{eq12:anse}
    \text{ANSE} =10\log_{10}  \bigg\{\frac{1}{K}\sum_{k=1}^{K}\frac{\mathbb{E}[e^2_k(n)]}{\mathbb{E}[d^2_k(n)]}\bigg\}.
\end{equation}
\begin{figure}[!t]
    \centering
    \includegraphics[width=0.9\columnwidth]{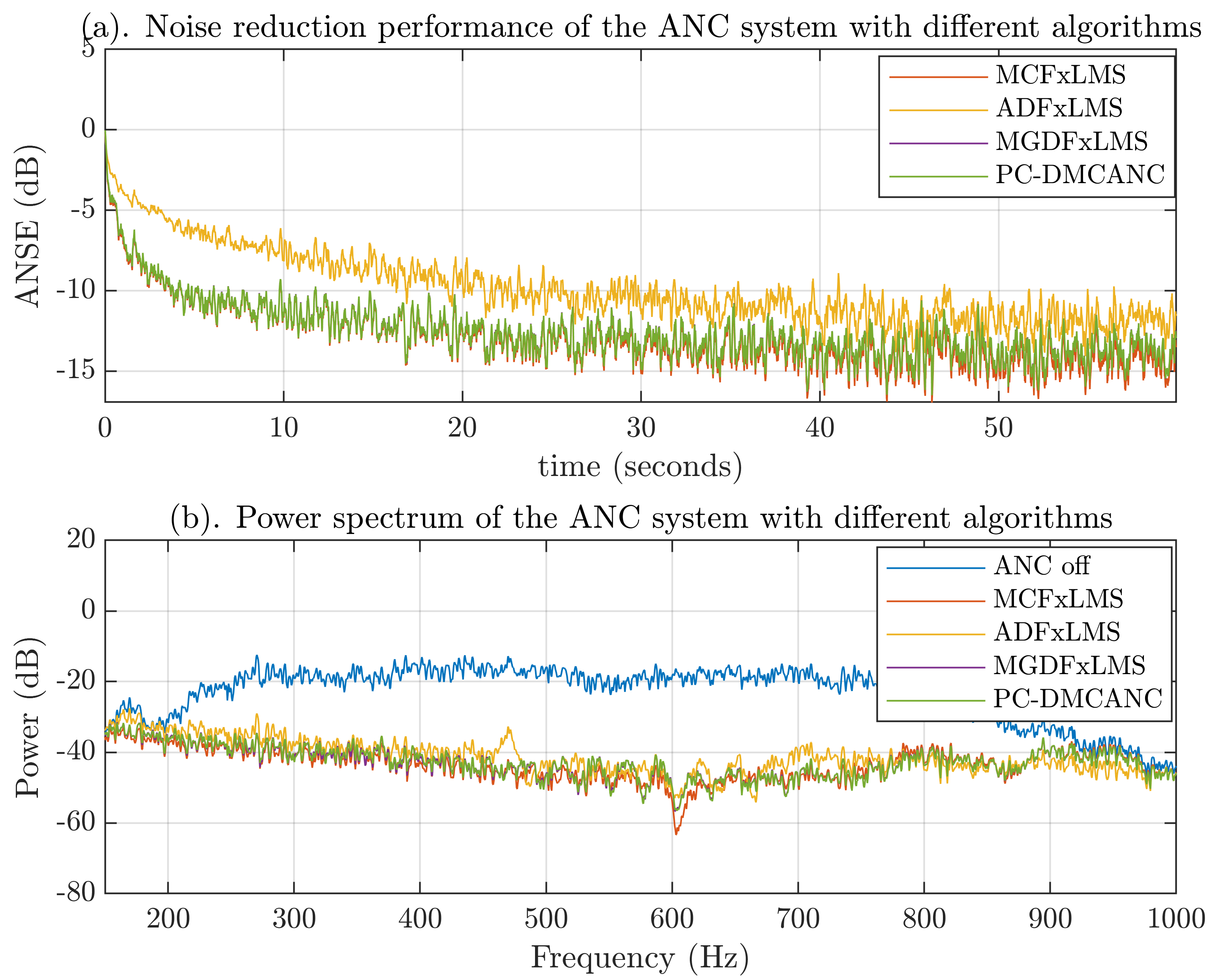}
    \caption{Noise reduction performance for broadband noise: (a) ANSE comparison of various MCANC systems; (b) Power spectrum comparison of various MCANC algorithms.}
    \label{fig:case1}
\end{figure}

\subsection{Noise reduction performance under an ideal distributed network}
\begin{figure}[!t]
    \centering
    \includegraphics[width=0.9\columnwidth]{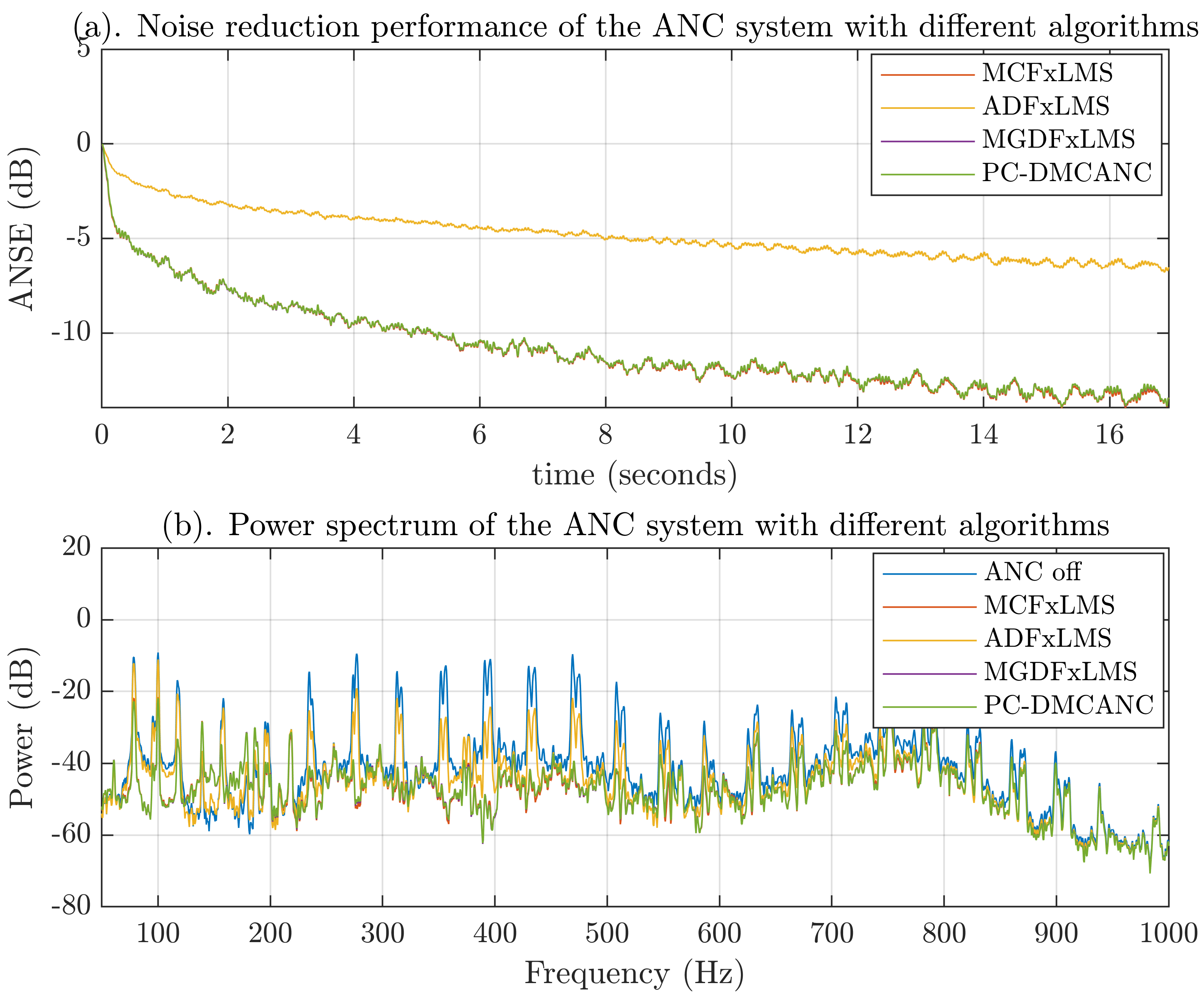}
    \caption{Noise reduction performance for real recorded compressor noise: (a) ANSE comparison of various MCANC systems; (b) Power spectrum comparison of various MCANC algorithms.}
    \label{fig:case31}
\end{figure}

The majority of the DMCANC algorithms are under the assumption of an ideal network. To validate the performance of the proposed algorithm, a broadband noise ranging from 200 to 900Hz is chosen as the primary noise, and the proposed PC-DMCANC is also compared with the conventional centralized MCFxLMS \cite{Elliott1987MEANC}, ADFxLMS \cite{Li2023Distributed}, and MGDFxLMS \cite{ji2025mixed}. The step size is selected as $5\times10^{-7}$ for all the algorithms. As illustrated in Fig.~\ref{fig:case1}, the proposed PC-DMCANC system can achieve almost identical performance to the centralized algorithm but outperforms the ADFxLMS algorithm. The power spectrum shown in Fig.~\ref{fig:case1}(b) further demonstrates the effectiveness of the proposed method on attenuating broadband noise. The simulation results using a real recorded compressor as primary noise are shown in Fig.~\ref{fig:case31}. It can be observed that the proposed PC-DMCANC method is applicable in practical scenarios, including real noise sources and acoustic paths.

\subsection{Noise reduction performance under communication delays}
\begin{figure}[!t]
    \centering
    \includegraphics[width=0.9\columnwidth]{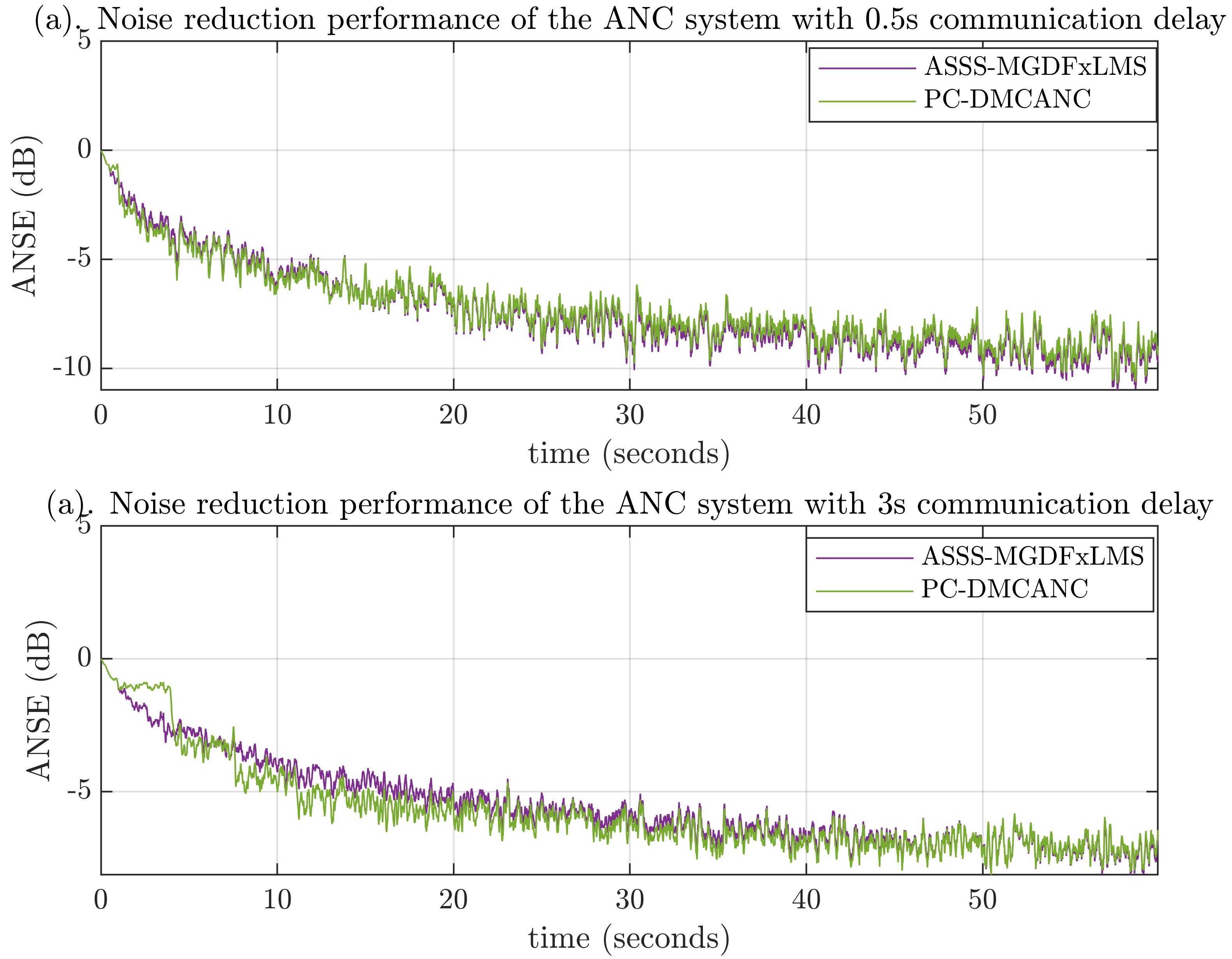}
    \caption{The noise reduction performance of different algorithms under different communication delays: (a) ANSE comparison with 0.5 seconds delay in the distributed network; (b) ANSE comparison with 3 seconds delay in the distributed network}
    \label{fig:case2}
\end{figure}

In practice, the distributed network may not be stable. Hence, in this simulation, we verify the robustness of the proposed PC-DMCANC by introducing communication delays into the system. The primary noise is a broadband noise, and ASSS-MGDFxLMS \cite{ji2025mixed} is selected for comparison. The initial step size for ASSS-MGDFxLMS and the step size for the PC-DMCANC are chosen as $10^{-7}$. From Fig.~\ref{fig:case2}, it can be seen that both the proposed PC-DMCANC and ASSS-MGDFxLMS algorithms have the same noise reduction effect under different communication delays. However, the PC-DMCANC approach requires less communication frequency than the ASSS-MGDFxLMS. 

It can be further observed from Fig.~\ref{fig:case2}(b) that the PC-DMCANC method has a flat region, where the system is operating as a fixed filter to maintain the noise reduction performance in the communication phase. However, it can still achieve almost identical performance as the ASSS-MGDFxLMS algorithm as time increases. Therefore, the proposed PC-DMCANC system can effectively deal with non-ideal networks and exhibits certain practical values.

\section{Conclusions}\label{sec:conclusion}
In this paper, a proactive communication distributed MCANC (PC-DMCANC) framework is proposed that combines adaptive–fixed‐filter switching with a mixed cumulative gradient (MCGD) combination strategy to address the twin challenges of communication overhead and system stability in large-scale ANC networks. Each node runs a local FxLMS algorithm and monitors its own residual noise, suspending adaptation and switching to a robust fixed filter whenever performance degrades; simultaneously initializing a communication request for exchanging cumulative gradients. Upon receiving other nodes' data, the node employs an MCGD method to update its control filter, subsequently reverting to the adaptive mode. Simulation results under both ideal and delayed network conditions demonstrate that PC-DMCANC achieves satisfactory noise reduction performance while dramatically reducing communication frequency and maintaining stability. Therefore, PC-DMCANC represents a robust, communication-efficient solution for real-world distributed ANC deployments. Future work will explore real-time validation in wireless sensor networks and a more efficient way to avoid mode switching in the system.

\section*{Acknowledgment}
This work was supported by the Ministry of Education, Singapore, through Academic Research Fund Tier 2 under Grant MOE-T2EP20221-0014 and Grant MOE-T2EP50122-0018.

\printbibliography

\end{document}